\documentclass[reprint,
superscriptaddress,
preprintnumbers,
 amsmath,amssymb,
 aps,
floatfix,
]{revtex4-2}

\usepackage{lipsum}
\makeatletter
\newcommand*{\balancecolsandclearpage}{%
  \close@column@grid
  \cleardoublepage
  \twocolumngrid
}
\makeatother

\newcommand{\oursection}[1]{\section{#1}}
\newcommand{\revision}[1]{#1}

\usepackage{hyperref}
\usepackage{xcolor}
\usepackage{graphicx}% Include figure files
\usepackage{dcolumn}% Align table columns on decimal point
\usepackage{bm}% bold math
\begin{document}

\preprint{SLAC-PUB-17562}

\title{Near-critical supernova outflows and their neutrino signatures}

\author{Alexander Friedland}
\affiliation{SLAC National Accelerator Laboratory, Stanford University, Menlo Park, CA 94025, USA}
\author{Payel Mukhopadhyay}
\affiliation{SLAC National Accelerator Laboratory, Stanford University, Menlo Park, CA 94025, USA}
\affiliation{Physics Department, Stanford University, Stanford, CA 94305, USA}

\date{August 27, 2022}

\begin{abstract}
We demonstrate that the neutrino-driven outflows inside  exploding core-collapse supernovae possess a special property of near-criticality, that is, they are on the edge of forming termination shocks. We derive a novel criterion for the formation of the shock, in terms of the fundamental parameters of the problem: the neutrino luminosity and energy as well as the properties of the protoneutron star. The criterion provides a unified description of the available numerical results and motivates future simulations.  The property of near-criticality makes the neutrino signatures of the termination shocks a sensitive diagnostic of the physical conditions around the PNS several seconds into the  explosion. The expected signal at DUNE is found to be statistically significant.

\end{abstract}

\maketitle

%\tableofcontents

\oursection{Introduction}
Core-collapse supernovae (CCSNe) have shaped our galaxy~\cite{Efstathiou:2000xp,Scannapieco:2008vf}, played an essential role in the formation of the solar system~\cite{CAMERON1977447,Banerjee:2016tzd}, and created many elements found on Earth, including oxygen~\cite{Thielemann:1996mc}. 
A CCSNe is triggered when the core of a massive star at the end of its evolution collapses to nuclear densities~\cite{Colgate:1960zz,Colgate:1966ax,Bethe_1985,Bethe:1990mw,Colgate:1993rz}, creating extreme physical conditions that cannot be reproduced in the laboratory.
Modeling of the explosion mechanism brings together considerations of nuclear and particle physics, general relativity, transport theory, and hydrodynamics. Modern 3D simulations of  CCSNe are some of the most sophisticated and resource-intensive supercomputing calculations in the world, e.g., ~\cite{Lentz:2015,Muller:2017hht,Radice:2017ykv,Vartanyan:2018xcd,Burrows:2019rtd}.

The focus here will be on the physical conditions in the region surrounding the protoneutron star (PNS), which is formed by the collapsed core. Seconds after the explosion is launched, the neutrino burst, which is observed to last ${\cal O} (10\mbox{ s})$~\cite{Kamiokande-II:1987idp,Bionta:1987qt} continues to heat the material in the so-called \emph{gain region}~\cite{Bethe:1992fq}, just above the PNS surface. The heating drives an outflow of matter, continuing to increase the energy of the explosion~\cite{Lentz:2015,Bollig:2020phc}. 
This outflow forms a low-density, high-entropy region around the PNS~\cite{Wilson:1982}, known as the hot bubble~\cite{Bethe:1990mw,Woosley:1992ek}.

The properties of the matter profile in the hot bubble are essential for two crucial applications. First, the outflow is a potential site for a number of nucleosynthetic processes~\cite{Qian:1996xt,Sumiyoshi:1999rh,Thompson:2001,Otsuki:1999,Wanajo:2001,Wanajo:2006rp,Arcones:2006, Wanajo:2010mc, Arcones:2012wj,Roberts:2010wh, Kuroda:2008, Arcones:2011zj,Bliss:2018}. In a companion paper~\cite{nu-p-paper}, we demonstrate that the properties of the outflows established in the present work play a crucial role in the success of the $\nu p$-process~\cite{Frohlich:2005ys,Wanajo:2006rp,Wanajo:2010mc} for realistic supernova conditions.
Second, the matter profile above the PNS affects neutrino flavor transformations. In particular, density features in and around the hot bubble, such as shocks and turbulence, can potentially have a significant impact on the detectable neutrino signal~\cite{Schirato:2002,Tomas:2004gr,Friedland:2006,Gava:2009pj,Lund:2013uta}. \revision{According to modern calculations of the neutrino fluxes (e.g., \cite{Bollig:2020phc}), these oscillation effects should be most pronounced in neutrinos, rather than in antineutrinos~\cite{paper2}.}

\revision{This points to a very attractive possibility that the neutrino signal at DUNE---which is uniquely sensitive to electron neutrinos at supernova energies---could be used to learn about the conditions around the PNS. The idea is to use this signal both as a diagnostic of the explosion mechanism and as a probe of the conditions for nucleosynthesis.}
To make this connection, however, requires a robust understanding of the outflow physics, the resulting density features in the hot bubble, and their neutrino signatures. The problem is inherently multi-disciplinary and a systematic study of several crucial aspects is lacking. Specifically, one needs to quantify the conditions of the formation of the \emph{termination shock} in the outflow and to explore its impact on neutrino oscillations.

In this letter, we develop a unified treatment of subsonic and transonic outflows and use it to derive a criterion for shock formation, in terms of the fundamental parameters of the problem. 
We show that the supernova outflows possess a special property, which we call \emph{near-criticality}. This property makes the neutrino signal a sensitive diagnostic of the explosion. 
Examples of the smoking-gun signatures of this physics at DUNE are presented and their statistical significance quantified. 

\begin{figure}[bt]
\centering
\includegraphics[width=0.96\columnwidth]{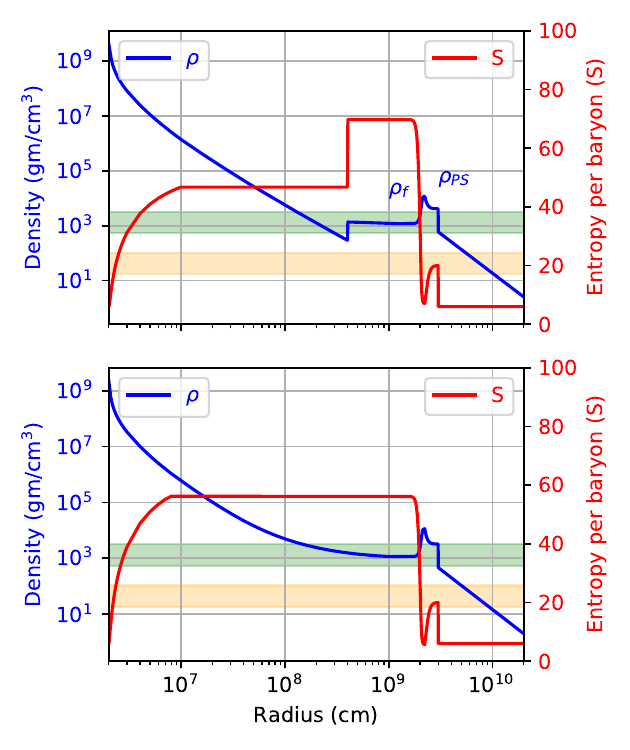}
\caption{\label{fig:cartoon} Two possible profile types several seconds into the explosion: with (\emph{top panel}) and without (\emph{bottom panel}) the outflow termination shock. Shown are density ($\rho$) and entropy per baryon ($S$). The horizontal color bands indicate the ranges of densities where the atmospheric (\emph{green}) and solar (\emph{orange}) MSW resonances occur.}
\end{figure}

\oursection{Termination shock: state of simulations}
While modeling neutrino signals, some investigations consider outflows with strong termination shocks~\cite{Tomas:2004gr,Lund:2013uta}, while others employ smooth outflow profiles~\cite{Schirato:2002}. %The former are called neutrino-driven ``winds'', while the latter are termed ``breezes''. 
%For studies of neutrino signals, the difference between the two is crucial, as we discuss below.
The two possibilities are illustrated in Fig.~\ref{fig:cartoon}. The neutrino signal predictions in the two scenarios can be significantly different, as we will see later. 

The diversity of the profiles used in the oscillation studies reflects the state of supernova simulations.
Termination shocks are prominent in, e.g.,~\cite{Tomas:2004gr,Arcones:2006}, while they are  absent in \cite{Woosley:1992ek,Witti:1994}, which informed the profiles in~\cite{Schirato:2002}. Ref.~\cite{Fischer:2009} is particularly noteworthy: while some models there do not have termination shocks, in others the shock appears episodically. Modern 3D simulations likewise do not obtain a unique answer. For instance, the outflow in \cite{Bollig:2020phc} is subsonic, while the simulation of a $9.6M_{\odot}$ progenitor in \cite{Stockinger:2020hse} features a prominent termination shock. 

\revision{What is one to make of these seemingly disparate simulation results? This turns out to be a very fruitful question to pursue. The key point here is that, whatever the physical criterion of the shock formation is, one would not a priori expect the conditions in a supernova to be tuned to be ``on the edge'' between the two regimes. Hence, one may be tempted to conclude that the outflows should be either always subsonic or always supersonic and to ascribe the different results in the literature to different numerical treatments. Yet, the alternative, however a priori unlikely, is worthy of exploration: if the system is indeed ``on the edge'', the formation of the termination shock sensitively depends on the details of the physical conditions and its observation in neutrinos could then serve as a sensitive diagnostic. To see which of these possibilities is correct, we need to establish a quantitative shock formation criterion.}

\oursection{Termination shock: a semi-analytical model}
To put our problem in perspective, it is useful to consider other astrophysical systems where outflows from a central object occur. A classic example is the solar wind~\cite{Parker:1963}; another one is an outflow in a pulsar wind nebula (PWN)~\cite{Slane:2017bje}. In both cases, the outflows form strong termination shocks when running into the surrounding interstellar media. The mass loss rates in these systems depend on the details of the heating rates near the central objects, but \emph{the existence of the shocks does not}. Evidently, the case of the supernova neutrino-driven outflow is in some aspects unique. How?

Importantly, in modeling the solar wind, the processes of wind acceleration and termination are treated independently. This is obviously justified, given the wind shock is observed at a distance of 94 astronomical units from the Sun~\cite{Stone:2005vj}, far away from the corona where the material is launched. This ``factorization" of the problem is so ubiquitous for stellar outflows that it was, in fact, implicitly adopted in the seminal paper on the neutrino-driven wind models~\cite{Duncan1986}. 
In view of the preceding discussion, however, it is worth to systematically examine when this separation is valid.

In our investigation, we follow the standard approach of the existing semi-analytical studies~\cite{Duncan1986,Qian:1996xt,Otsuki:1999, Wanajo:2001,Thompson:2001}, which assume spherical symmetry and a steady-state approximation. This framework allows us to distill the relationship between shock formation and the fundamental parameters of the system, which may not be easy to do with full simulations of the explosion. 

With this, the problem reduces to a system of ordinary differential equations, as summarized in Appendix~\ref{app1} (see \cite{paper2} for
full details).
Compared to the studies in~\cite{Duncan1986,Qian:1996xt,Otsuki:1999, Wanajo:2001,Thompson:2001}, we include in our equations the variation of the number of effective relativistic degrees of freedom, $g_\star$~\cite{paper2}. This is relevant because the material in the outflow starts out at the temperatures of several MeV near the PNS surface and cools to temperatures below the electron mass as it expands.

In addition to the governing equations, just as important is the correct identification of the boundary conditions. 
\revision{Rather than being a mathematical formality, this step goes to the core of the physics of the problem. 
The conditions at the starting radius are completely specified by two quantities: temperature, $T_i$, and entropy per baryon, $S_i$. However, these two initial conditions alone are not sufficient to fully specify the problem, since we have three equations for three unknown functions: $v$, $T$, and $S$. Sometimes, the mass loss rate $\dot{M}=4\pi r^2 v \rho$ is loosely called the third boundary conditions. However, physically, $\dot{M}$ is a \emph{derived} quantity.} The necessary third condition is provided by the pressure of the material surrounding the hot bubble, $P_f$. This outer boundary condition for the pressure is dependent on the progenitor profile, through the mass swept up by the expanding shock, as detailed in Appendix~\ref{app2}.

\revision{One may object here that the standard treatment of stellar winds does not explicitly invoke the outer boundary condition. It is, however, implicitly present---the solution is constructed under the assumption that the wind expands into vacuum, that is, $P_f=0$. This vacuum solution, in fact, plays an essential role in our construction described below.}

Thus, mathematically, we are dealing with a boundary value problem, rather than an initial value problem. Physically, \emph{the character of the outflow is determined by the interplay of the conditions near the PNS and those at the edge of the hot bubble.}

\revision{Importantly, we need to be able to solve the problem for any value of $P_f$. 
For this, we developed the following algorithm. First, we attempt to match $P_f$ by shooting the initial velocity, $v_i$. If a solution can be found, the process concludes and we obtain a smooth outflow that is \emph{always subsonic}~\cite{Qian:1996xt,Otsuki:1999}. For sufficiently small $P_f < P_f^{crit}$, however, no $v_i$ can be found in this way. Attempting a velocity higher than a certain critical value, $v^{crit}$, gives a physically untenable answer. 
The physics of this mathematical behavior is the formation of the shock. The pre-shock part of the outflow is now causally disconnected from the outside and hence is the same as in the case of the expansion in vacuum, mentioned above. Accordingly, once $v^{crit}$ is reached, the algorithm changes. We start with an auxiliary ``wind" solution which describes expansion into a vacuum. How to find it is well known since the classical solar wind work by Parker~\cite{Parker:1963}. The outflow in this case accelerates to supersonic speeds and the sonic point, being a mathematical singularity, requires some care~\cite{Duncan1986,Thompson:2001}. Next, one uses this wind solution up to the location of the termination shock, $R_{ts}$. The density and velocity jumps at $R_{ts}$ are given by the Rankine-Hugoniot jump condition,
 \begin{equation}
     v_2/v_1=\rho_1/\rho_2 = (2 T_1 S_1/m_N + v_1^2)/(7 v_1^2),
 \end{equation}
 where $v_1$, $T_1$, and $S_1$ are \emph{pre-shock quantities} and therefore are taken from the wind solution at point $R_{ts}$. The postshock solution is subsonic and is followed using the standard technique. Iterating $R_{ts}$ is sufficient to match any given value of $P_f < P_f^{crit}$. Thus, one uses the shooting method in both regimes, in the subsonic regime shooting $v_i$ and the supersonic, $R_{ts}$. }

\begin{figure}
\centering
\includegraphics[width=0.88\columnwidth]{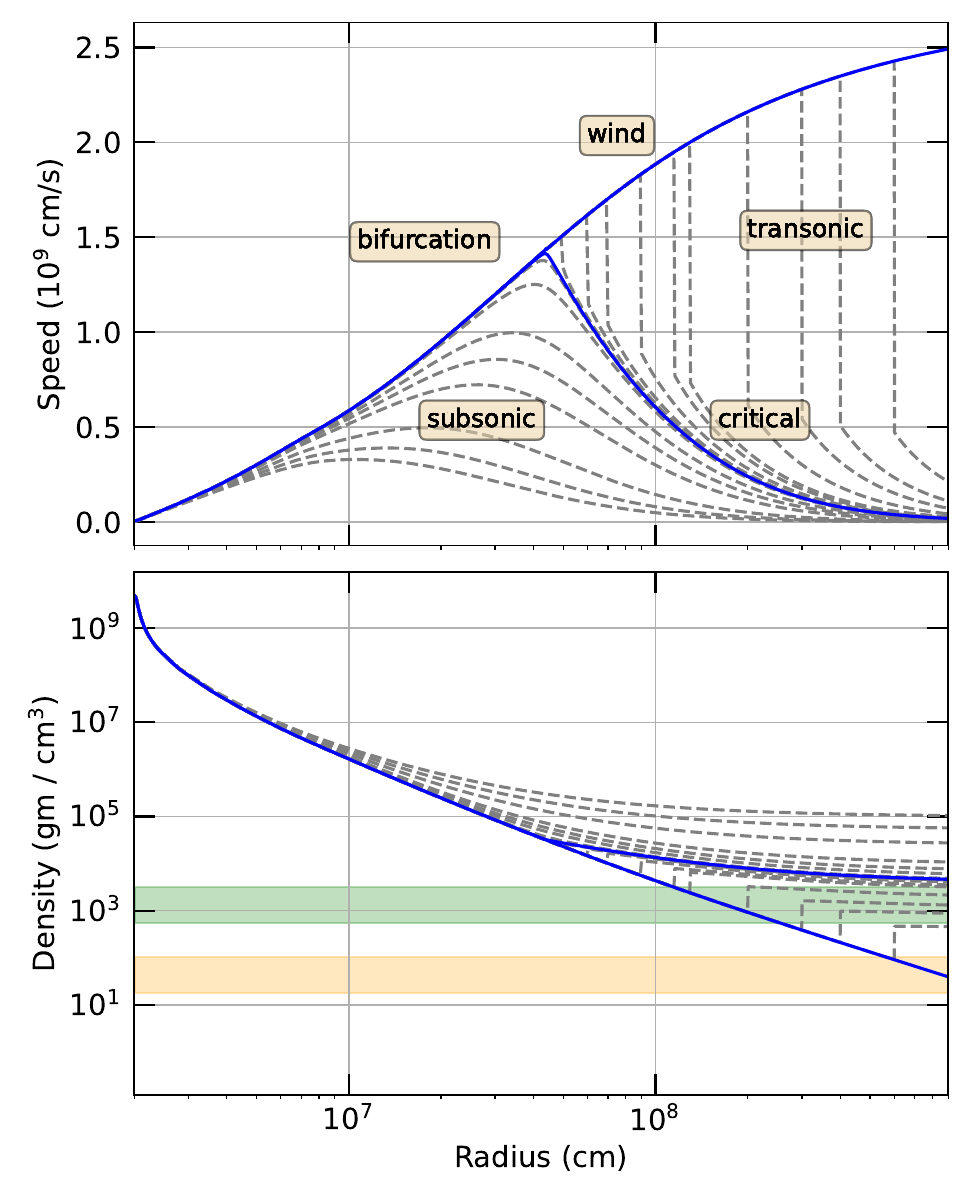}
\caption{\label{fig:4panels} Profiles of speed and density of the outflow as the far pressure $P_{f}$ is varied. Termination shocks start to form as $P_{f}$ falls below a critical value $P_{f,crit}$. The solid curves mark the wind solution (expansion into a vacuum) and the critical solution separating the subsonic and transonic regimes.}
\end{figure}

With this procedure, we obtain a family of solutions for a range of physically plausible values for the final pressure $P_f$, imposed at $10^4$ km. We fix the neutrino luminosities $L_{\nu_e}=L_{\bar\nu_{e}}=8 \times 10^{51}$ erg/s, the appropriate energy average $\epsilon \equiv \sqrt{\left<E_\nu^3\right>/\left<E_\nu\right>}=20$ MeV, the mass of the PNS, $M=1.4 M_\odot$, and the gain (starting) radius $R=20$ km~\cite{Fischer:2009}. The results are plotted in Fig.~\ref{fig:4panels}. Further details can be found in Appendices~\ref{app3} and \ref{app4}. 

\oursection{Critical flow}
\revision{Fig.~\ref{fig:4panels} illustrates the family of solutions obtains upon varying $P_f$. Both phases of the flow can be clearly seen.} For small $P_f$, the outflow accelerates past the speed of sound and then encounters a termination shock. Information from large radii travels only up to the shock location; the pre-shock part of the outflow ``believes" it is expanding into empty space and is thus universal. \revision{The region of neutrino energy deposition lies closer than the sonic point. Hence,} in this regime, the factorization between wind acceleration and termination is justified. In contrast, for large $P_f$, we obtain a family of smooth curves. These curves are subsonic and thus fully causally connected, with no factorization possible. 

The solid curve separating these two phases is a \emph{critical flow}. This curve can be viewed as a limiting subsonic flow, which develops a cusp singularity, or a limiting transonic flow which has a vanishing discontinuity in density and velocity (although a finite discontinuity in acceleration). The cusp singularity, seen at $r\sim500$ km, occurs at the sonic point, where the solution \emph{bifurcates}---the upper branch corresponds to the wind  expanding into a vacuum (also solid in the figure). The two branches overlap to the left of the sonic point and are characterized by the same mass loss rates and entropy per baryon values. 

Given the conditions at the gain radius and the neutrino heating rates, the critical flow is obtained when the far pressure is tuned to a certain specific value,  $P_{f,crit}$. An important physical analogy can be made to a system that can be studied in the laboratory: a flow of compressible gas through a nozzle~\cite{LandauLifshitzHydro,HydroTextbook}. While a nozzle flow has a completely different geometry and lacks gravity, it exhibits the same subsonic/transonic transition as the back pressure is adjusted. In particular, one can also tune the back pressure such that the nozzle flow develops a cusp singularity when the velocity profile touches the sound speed. \revision{In the language of nozzle hydrodynamics, this is the onset of the chocked flow regime.}

This brings us to a crucial point about outflows in a supernova. While in the laboratory one can fine-tune the conditions to critical, in astrophysical systems generically no such tuning is expected. Indeed, the stellar wind and PWN cases mentioned above are far in the transonic regime, with ample separation between acceleration and termination regions. Yet, in a supernova the situation is rather peculiar: during the first several seconds after the explosion, the region surrounding the hot bubble can have temperatures of order 0.1 MeV. This happens to give pressure values close to $P_{f,crit}$. The supernova outflows are thus, remarkably, \emph{near-critical}.

\revision{What is even more remarkable is the time evolution of the conditions in the flow. As the front shock expands, the density and pressure behind it drop. At the same time, so do neutrino luminosities driving the outflow. These two effects tend to partially cancel out, with the result that a flow close to critical remains so for several seconds as the explosion develops. In certain cases, the critical condition can even be crossed during the development of the explosion, resulting in intermittent termination shocks.}

These observations allow us to reconcile the seemingly disparate numerical results in the simulation literature. The formation of the shocks depends on the detailed conditions in the simulation, such as neutrino luminosities and the PNS parameters. In order to do this, we need to establish quantitative conditions for criticality. 

\oursection{Scaling law}
We investigated the dependence of the critical flow on the fundamental parameters of the problem: neutrino luminosity $L$ and energy $\epsilon$, and the PNS mass $M$, and radius $R$.
Since pressure in the outer layers of the hot bubble is radiation-dominated, one can speak of the critical temperature, $T_{f,crit}$. Moreover, 
given the value of entropy per baryon in these layers, $S_f$, one can speak of the critical matter density, $\rho_{f,crit} \sim T_{f,crit}^3/S_f$.
By numerically solving our model on a grid of points, we discovered that the critical values of these quantities follow the following power-law scaling relations: 
\begin{eqnarray}
T_{f,crit} &\simeq& (112\mbox{ keV}) L_{52}^{0.702} E_{\nu20}^{1.404} M_{1.4}^{-0.96} R_{20}^{0.08},
\label{eq:scaling_law_T}\\
\rho_{f,crit} &\simeq& 
(8.1 \times 10^3 \mbox{ g/cm$^3$})
L_{52}^{2.61} E_{\nu20}^{5.2} 
M_{1.4}^{-4.0} R_{20}^{1.03}.\;\;\;
\label{eq:scaling_law_rho}
\end{eqnarray}
Here $L_{52}=L/10^{52}$ erg, $E_{\nu20} = E_{\nu}/20$ MeV, $M_{1.4}=M/1.4 M_\odot$, and $R_{20}=R/20$ km. These relations are satisfied to very good accuracy, as shown in Supplemental Material in Appendix~\ref{app5}.

As will be explained in detail in \cite{paper2}, these scaling relations can be understood analytically. Namely,
\begin{eqnarray}
    T_{f,crit} 
    &\propto& L^{5/6} E_\nu^{5/3} M^{-1},\\
    \rho_{f,crit} 
    &\propto& L^{8/3} E_\nu^{16/3} R^{2/3}M^{-4}.
\end{eqnarray}
The agreement is quite satisfactory, especially in view of the fact that these relations are derived assuming constant $g_\star$, while the numerical results vary $g_\star$ with the temperature.

We have explicitly checked that Eq.~(\ref{eq:scaling_law_rho}) makes it possible to reconcile all the numerical results in the literature. For example, for the heating rates similar to Ref. \cite{Fischer:2009}, the boundary between progenitors that develop termination shocks at $t\sim2-3$ s and those that do not lies around 12$M_\odot$, consistent with the results in that paper. For higher heating rates, such as those assumed in Ref.~\cite{Arcones:2006}, robust shocks are obtained even for progenitors of $15M_\odot$. Still heavier progenitors, such as the $20 M_\odot$ model of \cite{Woosley:1992ek,Schirato:2002}, tend to give subsonic outflows, because of the high postshock densities. Nonetheless, we checked that by raising the $\nu_e$ energy to that of $\nu_x$ in the model of~\cite{Bollig:2020phc} brings even explosions of 18-19 M$_{\odot}$ progenitors close to shock formation. We see that, thanks to the near-criticality property, the existence of the shock is indeed sensitive to the detailed conditions in the explosion over a broad range of progenitor masses. 

\begin{figure*}[!htb]
\centering
\includegraphics[width = 0.49\linewidth]{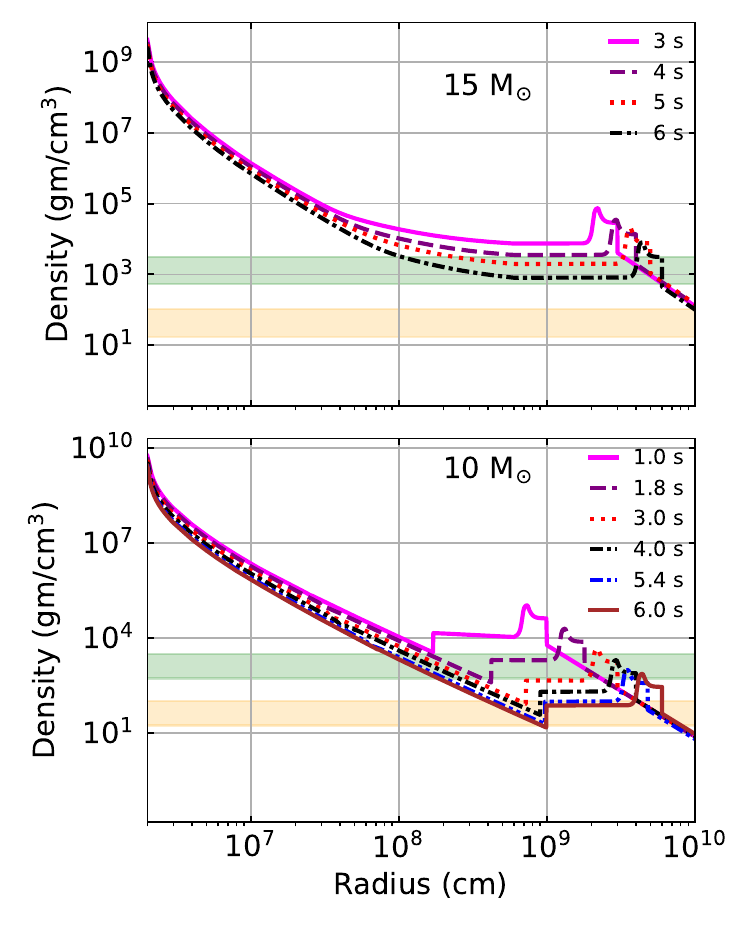}
\includegraphics[width = 0.5\linewidth]{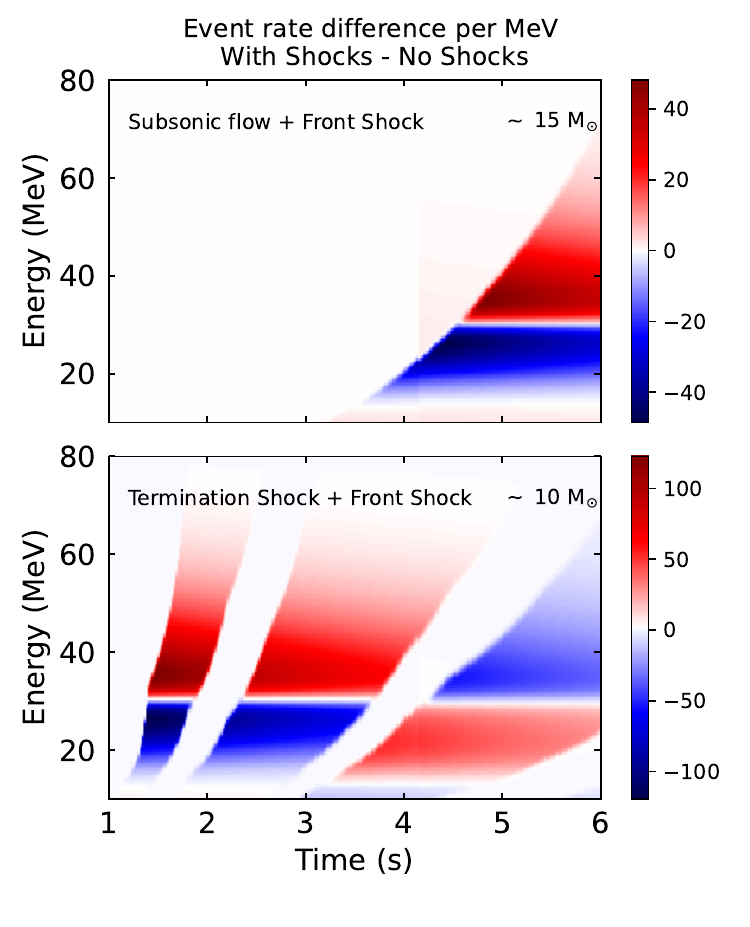} 
\caption{\label{fig:dunespectra} \revision{\textit{Left:} Time evolving outflow profiles for two illustrative explosion models: a $15 ~M_\odot$ progenitor with a subsonic outflow  (\textit{top panel}) and a $10 ~M_\odot$ progenitor with a wind termination shock (\textit{bottom panel}). \textit{Right:} Characteristic patterns of spectral modulations in the DUNE far detector for the two explosion models of the left panel. The heat map shows enhancement/depletion of the signal compared to the reference case of no shocks.}}
\label{fig:snapshots}
\end{figure*}

\oursection{Neutrino signatures}
The preceding discussion leads us to the last question: What are the signatures at DUNE and are they statistically significant? First, let us briefly discuss the mechanism by which the shocks imprint themselves onto the observed neutrino signal.
The essential ingredient lies in the physics of the MSW effect~\cite{Wolfenstein:1977,Mikheev:1986}, which is adiabatic only when the density scale height in the range of resonant densities is greater than the neutrino oscillation length. The density jump at the shock front makes the flavor evolution there maximally nonadiabatic. In addition to the outflow termination shock~\cite{Tomas:2004gr}, one should also consider the same effect on the expanding front shock~\cite{Schirato:2002}.

Because of the nuclear properties of $^{40}$Ar, DUNE is mostly sensitive to electron neutrinos, $\nu_e$~\cite{DUNE:2020zfm}. The observed $\nu_e$ spectrum is composed of the original $\nu_e$ and $\nu_{\mu,\tau}$ spectra, combined in proportions dictated by the oscillation probabilities. The effect of the shocks then is to alter the permutation pattern for these input spectra. 

Since the MSW resonance density depends inversely on the neutrino energy, and since both the front and termination shocks evolve from higher to lower densities, the modulation of the oscillation pattern starts at low energies and sweeps to high energy. This characteristic pattern can provide a smoking-gun signature of this effect, as it is unlikely to be faked by the neutrino-decoupling physics at the PNS surface. 

\revision{The outflow profiles and modulation patterns for two illustrative progenitors  are shown in Fig.~\ref{fig:dunespectra}. For each time snapshot, we solve for the density and velocity profiles in the steady-state approximation given the instantaneous postshock pressure and neutrino fluxes (cf. \cite{Qian:1996xt,Otsuki:1999}). We repeat this process for a number of closely-spaced snapshots in a time window $t \sim 1-6$ s (see details in appendix 2). The resulting sequence of density profiles for 10 M$_{\odot}$ and 15 M$_{\odot}$ progenitors, found using this time-dependent quasi-steady-state (TDQSS) is shown in the left panels of Fig. \ref{fig:dunespectra}. A similar time dependent quasi-steady-state framework has also been described recently in \cite{Xiong:2020ntn}.
}

\revision{The corresponding signatures in DUNE are shown in the right panels. The average neutrino energies in our model are kept fixed at $\left<E_{\nu_e}\right>$ = 10.5 MeV, $\left<E_{\bar{\nu_e}}\right>$ = 15.5 MeV and $\left<E_{\nu_{\chi}}\right>$ =  17.7 MeV and the spectral pinching parameters used are $\alpha_{\nu_e}$ = 4.1, $\alpha_{\bar{\nu}_e}$ = 3.9 and $\alpha_{\nu_{\chi}}$ = 2.9. These values are taken from \cite{Keil:2002in}. Numerical simulations show that average energies and pinching evolve slowly during the time window of interest \cite{Bollig:2020phc}, therefore keeping these parameters fixed over the time window is a reasonable simplification. Exponentially falling luminosity profiles, $L_{\nu} = L_0 \exp{(-{t}/{t_e})}$, with $t_e = 2.5$ s are assumed. The luminosity evolution is taken to mimic the values presented in 1D numerical simulations of \cite{Arcones:2006}. The top panel shows the case of a $15 M_\odot$ progenitor, in which the outflow is fully subsonic throughout the 6 s time window. The only modulation effect in the 15 M$_{\odot}$ case comes from the expanding front shock, which arrives to the resonance layer only at $\sim4$ s. The bottom panel shows the case of a $10 M_\odot$ progenitor, where both the termination and the front shock are present. In this case, the termination shock starts to modulate the signal as early as $1.4$ s into the explosion. The rest of the signal is then dictated by the interplay of the effects of the two shocks.
}

\revision{We note parenthetically that the density feature at the edge of the hot bubble,  known as the \textit{contact discontinuity}, is treated here in the adiabatic approximation for neutrino oscillations following multi-D simulations that find that this discontinuity may
be washed out partially or completely due to instabilities \cite{Tomas:2004gr}. Possible effects of turbulent density fluctuations in this region deserve a separate exploration.}

The calculation assumes a DUNE-like 40 kt liquid argon detector, with parameters from \cite{Arcones:2006,Huedepohl:2009wh}. The oscillation calculation assumes the normal mass hierarchy, as seems to be favored by the latest oscillation data from T2K \cite{Abe:2017vif, Dunne:2020} and NOvA experiments \cite{Acero:2019ksn, Himmel:2020}, as reflected in global fits \cite{Esteban:2018azc,salas:2020}.  The signal is simulated using the SNOwGLoBES \cite{Snowglobes:2018} package, assuming a DUNE-like 40 kt liquid argon detector.

In addition to the MSW matter effects, the calculation includes collective neutrino oscillations~\cite{Duan:2006an}. The latter are computed in a 3-flavor, multi-energy, multi-angle, spherically symmetric approach (similar to \cite{Duan:2010bf}). The choice of the neutrino spectra~\cite{Keil:2002in} creates a characteristic split feature~\cite{Duan:2006jv} at the high energy of $\sim30$ MeV~\cite{Dasgupta:2009mg,Friedland:2010sc}. As seen in the figure, the split feature can actually be revealed by the passage of the shocks. 

Further details on the model input, oscillation physics, and analysis methods will be presented in~\cite{paper2}. As will be explained in that paper, for a typical galactic distance of 10 kpc, and given sufficiently good energy resolution, the modulation signal can have a $5\sigma$ significance. Thus, the signal of the shock not only has a unique character, but is well within the reach of observability.

\oursection{Conclusions}
We demonstrated that the neutrino-driven outflows inside  exploding core-collapse supernovae possess a special property of near-criticality, which makes the neutrino signatures of the termination shocks a sensitive diagnostic of the physical conditions around the PNS several seconds into the  explosion. While the shockwave and the central engine are shrouded from the outside by the stellar envelope, pronounced signatures of the termination shock are imprinted in the neutrino signal. These signatures open up additional physics opportunities for the upcoming large detectors, such as DUNE. 

Additionally, the outflow properties established here provide essential input into our analysis of the yields of $\nu p$-process nucleosynthesis. The conditions in the subsonic regime 1--2 seconds after the shock is launched turn out to be optimal, as described in~\cite{nu-p-paper}.
Beyond this, we hope that our findings can be applied to other environments,  such as binary mergers \cite{Metzger:2008av,Wanajo:2011vy,Perego:2014fma} or very massive stars~\cite{Fujibayashi:2015rma}.

\begin{acknowledgments}
We would like to thank the Institute for Nuclear Theory and the Kavli Institute for Theoretical Physics for their hospitality. We have greatly benefited from conversations with Huaiyu Duan, Frank Timmes, Gail McLaughlin and Rebecca Surman. We thank Adam Burrows and Thomas Janka for guidance with numerical simulations, Amanda Weinstein for discussions of astrophysical shocks, Kate Scholberg for help with SNOwGLoBES, and the members of the SLAC Theory Group for their continuing support. Both authors are supported by the U.S. Department of Energy under Contract No. DE-AC02-76SF00515.
\end{acknowledgments}

\bibliography{bibliography}% Produces the bibliography via BibTeX. 

\balancecolsandclearpage
\onecolumngrid
\appendix

\section{Modeling of the outflow}
\label{app1}

Our modeling is based on a system of ordinary differential equations describing a spherically symmetric, stationary flow for a compressible fluid whose pressure is dominated by radiation and mass is dominated by baryons. 
These equations can be obtained from three basic considerations: (i) Newton's second law for pressure and gravitational forces, (ii) the continuity equation for the flow of matter, (iii) the continuity equation for the flow of entropy. 
Approximating the configuration as a steady-state flow and assuming spherical symmetry, we have
\begin{eqnarray}
    \rho v \frac{dv}{dr} &=& -\frac{dP}{dr} - \frac{G M \rho}{r^2},
    \label{eq:fluidacceleration}\\
     \frac{d[r^2 \rho v]}{dr} &=& 0,
     \label{eq:massconservation}\\
     \frac{d[r^2 \rho v S /m_N]}{dr} &=& \frac{r^2 \rho \dot{q}}{T}.
     \label{eq:entropyconservation}
\end{eqnarray}
Here the dependent functions are the outflow velocity $v$, density $\rho$, and temperature $T$. Equivalently, one can choose $v$, $\rho$ and entropy per baryon $S= (2\pi^2/45) g_\star T^3 m_N/\rho$. The quantity $g_\star$ denotes, as usual, the effective number of relativistic degrees of freedom. For $T\gg0.5$ MeV, when the radiation consists of photons and relativistic electron-positron pairs, we have $g_\star\rightarrow5.5$, while for $T\ll0.5$ MeV only photons contribute and $g_\star\rightarrow2$. Pressure is a function of temperature ($P=g_\star \pi^2 T^4/90$) or of $S$ and $\rho$. 

\revision{Note that we are assuming a radiation-dominated atmosphere, which is known to be a good approximation~\cite{Qian:1996xt}. We briefly reproduce the argument here. The spectra of streaming neutrinos are formed at the neutrinosphere, which is characterized by densities $\rho_{\nu}\sim10^{11}$ g/cm$^3$, temperatures  $T_\nu\sim5$ MeV, and a radius of $R_\nu\sim20$ km. For a PNS mass of $M\sim1.4M_\odot$, the gravitational potential at $R_\nu$ is $GM/R_\nu \sim0.1$, in natural units. At the neutrinosphere, the pressure of the nucleon gas, $\sim \rho T_{\nu}/m_N$, dominates over that of radiation, $\sim T_{\nu}^4$. Due to the high surface gravity, the density just above $R_\nu$ drops precipitously with the radius $r$, with the Boltzmann scale height of  $R_{\nu} (T/m_{N}) (G M/R_{\nu})^{-1}\sim 1$ km. Within $\sim7$ km, density decreases by three orders of magnitude, while the temperature, determined by neutrino absorption and reemission, changes much more slowly. As a result, the pressure in the outer layers is dominated by radiation. The pressure and specific internal energy of the material become dominated by contributions from the relativistic particles when S $\sim$ 4. Most of the mass in this radiation-dominated layer resides close to the surface of the PNS, with $\sim$80\% located within $r \lesssim 50$ km. This is where most of the neutrino energy deposition takes place and we can view this shell as the engine of the outflow. The bottom of this radiation-dominated shell is defined by the balance between the neutrino heating and cooling. This region, known as the \emph{gain radius}, lies closely to the matter-radiation equality. Above it, pressure quickly begins to be dominated by radiation.}

\revision{The specific energy deposition rate $\dot{q}$ has two main contributions: ($e N$) cooling due to neutrino emission in $e^-/e^+$ absorption on nucleons ($\dot{q}_{eN} \propto T^6$); and ($\nu N$) heating due to neutrino/antineutrino absorption on nucleons in the medium ($\dot{q}_{\nu N} \propto L_{\nu} \epsilon_{\nu}^2$). 
%The heating term $\dot{q_{\nu N}}$ falls as $1/r^2$ as a result of neutrino flux dilution while the cooling term $\dot{q_{eN}}$ falls like $T^6$. The cooling term falls more quickly than the heating term and the above mentioned `gain layer' is established inside $r \sim$ 100 km where neutrino heating dominates cooling. 
 Here, $\epsilon_{\nu} = (\left<E_{\nu}^3\right>/\left<E_{\nu}\right>)^{1/2}$, where $\left<E_{\nu}^n\right>$ denotes the $n^{th}$ moment of the neutrino energy distribution. The first moment $\left<E_{\nu}\right>$ is simply the average energy of the neutrino distribution which we denote by $E_{avg,\nu} \equiv \left<E_{\nu}\right>$. Additionally, we also include cooling due to annihilation of electron-positron pairs into $\nu_e \bar{\nu}_e$, $\nu_{\mu} \bar{\nu}_{\mu}$ and $\nu_{\tau} \bar{\nu}_{\tau}$. This cooling goes like $\dot{q}_{e^+e^-} \propto T^9/\rho$ \cite{Qian:1996xt}. 
}

Equations (\ref{eq:fluidacceleration})-(\ref{eq:entropyconservation}) can be recast in the form:
\begin{eqnarray}
    \left(v  -  \frac{v_s^2}{v} \right)\frac{dv}{dr} &=&  \frac{2 v_s^2}{r}  - \frac{G M}{r^2} - \beta\frac{\dot{q}}{v},
    \label{eq:fluidacceleration2_full} \\
\dot{q} &=& v\frac{d}{dr}\left(\frac{v^2}{2} + 3 v_s^2 - \frac{GM}{r}\right),
    \label{eq:dIdr} \\
    v\frac{dS}{dr} &=& \frac{\dot{q}m_N}{T}.
    \label{eq:qdotdS}
\end{eqnarray}
Here, $v$ is the outflow velocity, $S$ is entropy per baryon and $\dot{q}$ is a specific energy deposition rate.  
Furthermore, 
\begin{eqnarray}
v_s^2 &=& \frac{ST}{4 m_N}\left(1+ \left( 3+\frac{d \ln g_\star}{d \ln T}\right)^{-1}\right),\\
\beta &=& \frac{1}{4}\left(1+ \left( 3+\frac{d \ln g_\star}{d \ln T}\right)^{-1}\right).
\end{eqnarray}
These expressions take into account the change in the effective number of relativistic degrees of freedom $g_\star$, which occurs as the outflow cools through the temperature of $e^+e^-$ annihilation between the PNS surface and the outer edges of the hot bubble. If the dependence of $g_\star$ on $T$ is neglected, Eqs.~(\ref{eq:fluidacceleration2_full})--(\ref{eq:qdotdS}) reduce to those in~\cite{Duncan1986,Qian:1996xt}.

\newpage

\section{Modeling of the far boundary condition} 
\label{app2}

\revision{
We have argued in the main text that to completely specify the outflow requires imposing a pressure boundary conditions at large radii. In this appendix, we present the details of this process and relate the pressure to the progenitor model. 
}

\revision{
Specifically, the ``far" pressure refers to the conditions in the outer parts of the hot bubble, post termination shock, and the surrounding lower-entropy medium. We stress that there is no discontinuity of pressure across the so-called contact discontinuity. The discontinuity in density seen Fig.~\ref{fig:cartoon} between $\rho_{PS}$ and $\rho_f$ comes entirely from the jump of entropy per baryon. The entropy inside the hot bubble is provided by the outflow and originates from neutrino heating near the PNS surface and also from the termination shock if the latter is present. The entropy per baryon in the surrounding region is lower that in the outflow, only $S\sim6$. Equality of pressures then implies that the density jumps by a ratio of entropies: $\rho_{PS}/\rho_f=S_{f}/S_{PS}\sim 10$. 
}

\revision{
Note that the equality of pressures across the contact discontinuity implies an approximate equality of temperatures, in the approximation of radiation pressure domination. This is seen to be well satisfied in numerical simulations (e.g., \cite{Fischer:2009}).
}

\revision{
A prominent contact discontinuity is seen in all 1-D simulations \cite{Fischer:2009,Arcones:2006}. In multi-D, this discontinuity may be washed out partially or completely due to mixing driven by hydrodynamic instabilities \cite{Tomas:2004gr,Wongwathanarat:2012,Stockinger:2020hse}. In general, the region of the contact discontinuity and the postshock ejecta layer can be turbulent \cite{Stockinger:2020hse}. Turbulent density fluctuations will have distinct neutrino signatures than front and termination shocks \cite{Friedland:2006} and a detailed treatment of turbulence is left for a future study. 
}

\revision{
The density $\rho_{PS}$ is in turn related to the mass of the ejected and the swept-up material inside the expanding front shock. Hence, it is dictated by the structure of the progenitor and the dynamics of the explosion. Let us briefly outline the relevant physical considerations.
}

\revision{
We can estimate the mass ejected at the shock revival stage using general physics arguments. Let us consider the early explosion stage, when the front is stalled at a typical distance of $R_{s}\sim200$ km. The typical temperature in the volume behind the shock at this time, $T_s$, is found by comparing absorbed and re-emitted neutrino energy. Taking into account the geometrical factors, one finds $T_{s}\sim T_\nu (R_\nu/2R_s)^{1/3} \sim 2.5$ MeV. Here the neutrinosphere is assumed to have $T_\nu\sim4$ MeV and $R_\nu\sim100$ km. The radiation energy stored in this region is $\sim 5\times10^{50}$ erg, which equals the gravitational binding energy of a $\sim0.03 M_\odot$ shell to a $1.4 M_\odot$ PNS. Thus, the initially launched material comprises only a few percent of the solar mass.
}

\revision{
As the front shock plows through the progenitor profile, this mass grows. How much mass accumulates by the time the shock reaches a radius $R_s(t)$ depends on the progenitor profile and the value of the \emph{mass cut} (the position of the mass element that separates ejected and accreted material). The main factor dictating the density evolution is simply dilution due to the expanding front shock~\cite{Woosley:1992ek,Roberts:2010wh}, 
\begin{equation}
    \rho_{PS} \sim M_{plow}/[(4 \pi/3)R_s^3], R_s = v_s t
    \label{eq:density_evol_postshock}
\end{equation}
where $M_{plow}$ is the mass plowed up by the front shock in the explosion and $R_s(t)$ is the radius of the front shock. $M_{plow}$ is a strictly progenitor dependent quantity because it depends on the mass cut. 
}

\revision{
For example, for a 10 M$_{\odot}$ progenitor, integrating a sample progenitor profile in \cite{Sukhbold:2015} up to $R_s(t) = $ 10,000 km  yields the swept-up mass $M_{plow}\sim 0.2 M_\odot$. An additional $\sim0.1 M_\odot$ of mass will be added between 10,000 km and 100,000 km. 
The front shock speed $v_s$ is taken to be 10,000 km/s and $t$ is time in seconds. Therefore, for a 10 M$_{\odot}$ progenitor, the density dilution roughly goes like, $\rho_{PS} \sim 0.2~M_{\odot}/[(4\pi/3) (10,000 ~ km/s~ \times t_{sec})^3] \sim 10^5/t^3 ~\mathrm{g ~cm^{-3}}$. This evolution in post-shock density for a 10 M$_{\odot}$ progenitor is consistent with numerical simulations \cite{Fischer:2009}.  
}

\revision{
We thus obtain a physical model of the far pressure as a function of time, for progenitors of different masses. This information is used as an input into our model to create a set of approximate steady-state solutions for the outflow in a time window $t \sim 1-6$ s, as discussed in the main text. Such steady-state solutions are found to be in good agreement with detailed numerical simulations in the literature, for example, \cite{Arcones:2006,Fischer:2009}. The physics behind this agreement is that the time it takes for the outflow material to reach the termination shock is $\mathcal{O}(10^{-1})$ s, while the explosion conditions refer on the second time scale.
}

\newpage

\section{Outflow profiles} 
\label{app3}
\begin{figure} [!htb]
\centering
\includegraphics[width=0.5\columnwidth]{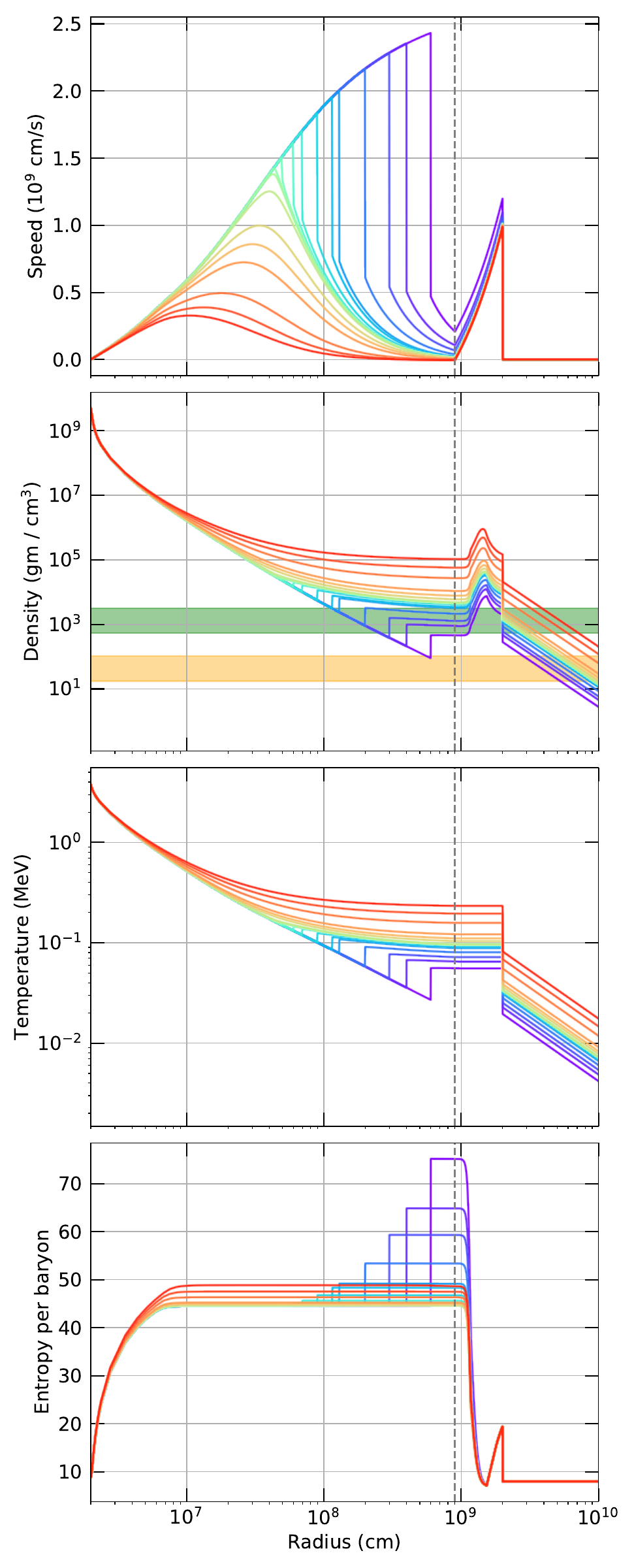}
\caption{\label{fig:4panels_app} Formation of the termination shock as a function of far boundary pressure. The four panels show the evolution of speed, entropy, temperature and density of the outflow as the far pressure boundary condition is varied. It is apparent that there exists a $P_{f,crit}$ below which termination shocks start to form.}
\end{figure}

Fig.~\ref{fig:4panels_app} shows an extended version of Fig. \ref{fig:4panels} in the main text, with the four panels depicting the profiles of the speed ($v$), density ($\rho$), temperature ($T$) and entropy ($S$) of the neutrino-driven outflows as the far boundary condition $T_f$ is varied. For completeness, the input parameters used for the simulation are $L_{\nu_e}$ = 8 $\times$ 10$^{51}$ erg/s, $\epsilon$ = 20 MeV, $M$ = 1.4 M$_{\odot}$ and $R$ = 20 km. 

The first and second panels are the same ones as in Fig.~\ref{fig:4panels}. 

The third panel shows the evolution of temperature ($T$) for both subsonic flows and flows with termination shocks. Since the pressure is dominated by radiation, the temperature is also a proxy for pressure, $P\propto T^4$. This panel thus illustrates pressure matching at our prescribed far boundary of $\sim$ 10$^4$ km to the pressure of the ``bump'' indicated with the vertical dashed line. The existence of the critical temperature ($T_{crit}$), separating breeze and wind solutions is clearly visible. 

The fourth panel shows the evolution of $S$. Additional entropy generation at the termination shock is clearly visible. This entropy generation is expected at termination shocks. Its possible impact on the r-process nucleosynthesis has been investigated in \cite{Arcones:2006}.

Comparison of the second and third panels shows that the density profile has additional features compared to the temperature profile, indicating the presence of layers with different entropies per baryon. The boundary condition for density $\rho_f$ can be obtained from far boundary pressure ($\propto T^4$) by assuming $S$ $\sim$ 50.

For the lowest far density considered in the second panel of Fig.~\ref{fig:4panels_app}, $\sim$ 450 g/cm$^3$, the termination shock is located at $\sim$ 6000 km, and the density jump is $\rho_2/\rho_1\simeq 5$. For vanishingly small far pressures, the shock radius will grow without bound, while the density jump reaches its limiting value for relativistic shocks, $(\gamma+1)/(\gamma-1)=7$, with the polytropic index $\gamma=4/3$. This shows that in our simulations, our termination shocks are not saturated in the strength of density jump which is another demonstration that the outflows are near-critical and not in either very subsonic or very strongly shocked regime.
 
The mass of the progenitor for which the critical curve is achieved is roughly estimated to be $\sim$ 11 M$_{\odot}$ in this calculation. The exact value of the progenitor mass above which termination shocks will not form will depend on $L$, $\epsilon$, $M$ and $R$. For higher luminosities, we have checked that termination shocks can form in higher mass progenitors of $M \sim 15 M_{\odot}$ as has already been seen in numerical simulations \cite{Arcones:2006}. 

\newpage

\section{Outflow parameters}
\label{app4}

Table \ref{table:lookup shock} shows the values of neutrino outflow parameters as the far boundary condition, $P_f$, is varied and $L$, $\epsilon$, $M$, and $R$ are held fixed to their values given in the main text and Appendix 2. Values of the final density, $\rho_f$, entropy $S_f$ and density jump across the shock, $\rho_2/\rho_1$ are also presented. It can be seen that the first six rows do not have any entries for the termination shock radius ($R_{ts}$) and the Shock jump factor ($\rho_2 / \rho_1$). This is because, in all these cases, the flow is subsonic. The critical solution has density $\rho_f$ $\simeq$ 4.5 $\times$ 10$^3$ g/cm$^3$, temperature $T \simeq 0.097$ MeV and entropy $S_f \simeq 45$. Note that this density $\rho_f$ is the density at the far boundary radius $\sim$ 10$^4$ km before the contact discontinuity in the high entropy region. 

Note that, strictly speaking, the critical parameter is temperature, which is directly related to the far pressure boundary condition. It can be converted to critical density assuming the typical values of $S\sim50$, as done for brevity in the main text. The critical temperature depends on the parameters of the problem as shown in the main text ({\it cf.} Eq.~\ref{eq:scaling_law_T}):

\begin{equation}
T_{crit} \simeq 112 \mbox{ keV}
\left(\frac{L}{10^{52} \mbox{ erg/s}}\right)^{0.702} \left(\frac{\epsilon}{20 \mbox{ MeV}}\right)^{1.404} \left(\frac{M}{1.4 M_{\odot}}\right)^{-0.96} \left(\frac{R}{20 \mbox{ km}}\right)^{0.08}.
\label{eq:powerlaw_T}
\end{equation}

\begin{table} [!htb]
\begin{center}
 \begin{tabular}{||c c c c c c||} 
 \hline
  & $T_{f}$ (MeV) & $R_{ts}$(10$^2$ km) & $\rho_{f}$ (10$^3$ g/cm$^3$) & $S_{f}$ & Shock jump factor ($\rho_2/\rho_1$)\\ [0.5ex] 
  \hline
 1 & 0.23 & -- & 104.02 & 48.4 & -- \\
\hline
 2 & 0.19 & -- & 56.50 & 47.4  & -- \\
\hline
 3 & 0.15 & -- & 27.01 & 46.3 & -- \\
\hline
 4 & 0.12 & -- & 10.61 & 45.2 & -- \\
\hline
5 & 0.11 & -- & 7.59 & 45.1 & -- \\
\hline
6 & 0.0985 & -- & 4.7 & 45.0 & -- \\
\hline
7 & 0.097 & -- & 4.50 & 44.8 & -- \\
\hline
8 & 0.096 & 5.0 & 4.4 & 45.0 & 1.12 \\
  \hline
9 & 0.094 & 7.0 & 4.37 & 45.5 & 1.63 \\
  \hline
10 & 0.093 & 9.0 & 3.98 & 46.7 & 2.06 \\
  \hline
11 & 0.090 & 11.5 & 3.48 & 48.2 & 2.5\\
  \hline
12 & 0.088 & 13.0 & 3.14 & 49.0 & 2.73\\
 \hline
13 & 0.080 & 20.0  & 2.09 & 53.0 & 3.5\\
 \hline
14 & 0.07 & 30.0  & 1.28 & 59.0 & 4.19\\
 \hline
15 & 0.064 & 40.0  & 0.85 & 64.8 & 4.6\\ 
 \hline
16 & 0.055 & 60.0  & 0.45 & 75.0 & 5.13\\
  \hline

\end{tabular}
\end{center}
\caption{Table of values for the termination shock radius $R_{ts}$ vs. far-end temperature ($T_{f}$) boundary conditions. Final density ($\rho_f$), entropy per baryon ($S_f$) and the shock jump factor ($\rho_2/\rho_1$) are also shown. The following parameters are held fixed: $L_{\bar{\nu}_e}$ = $8\times10^{51}$ erg/s, $\epsilon$ = 20 MeV, $R$ = 20 km, $M$ = 1.4 $M_{\odot}$. Empty entries imply that for those sets  of boundary conditions, there is no termination shock formation. }
\label{table:lookup shock}
\end{table}

\newpage

\section{Critical conditions: numerical results}
\label{app5}

\begin{figure} [!htb]
\centering
\includegraphics[width=0.7\columnwidth]{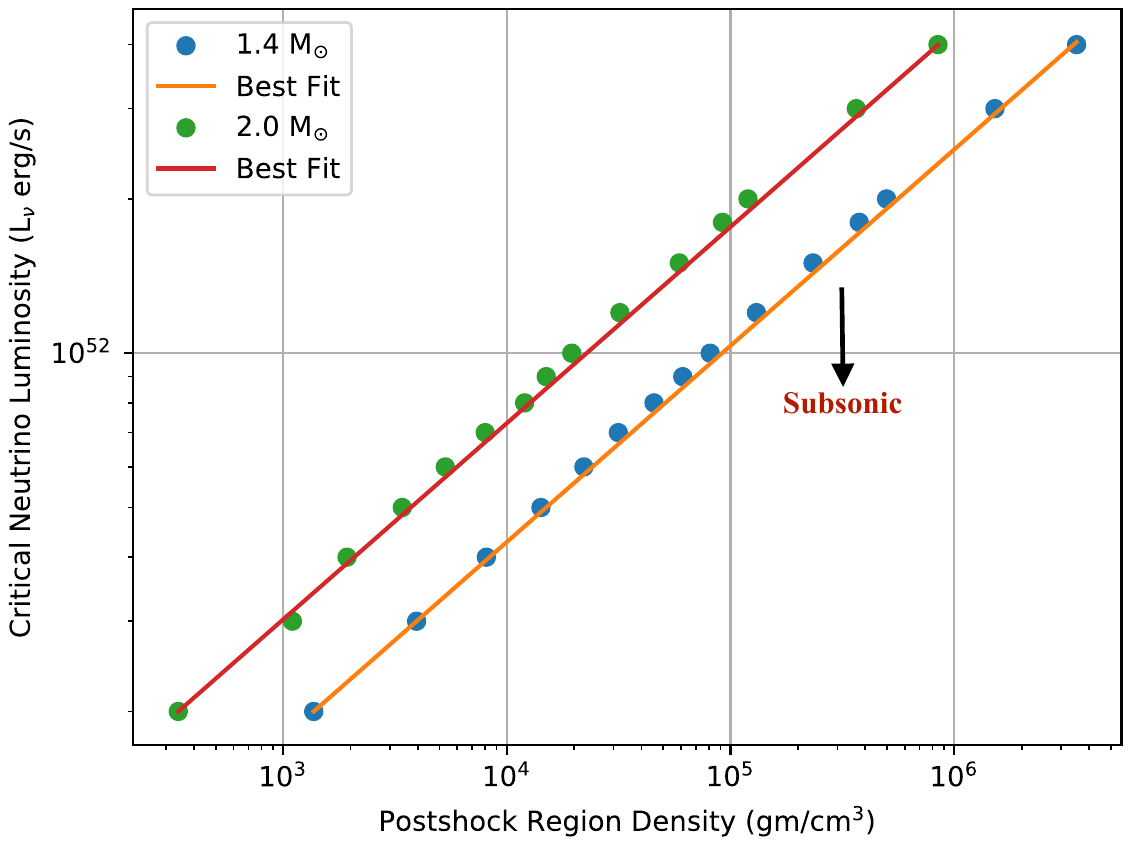}
\caption{Critical luminosity vs. far end postshock density ($\rho_{PS}$). Below the critical values, the outflow solutions are subsonic while above it they are supersonic and form termination shocks. The points show numerically determined critical parameters and the lines show the inferred power law fit. The two sets of points illustrate the dependence on the PNS mass values. }
\label{fig:critical_numerical}
\end{figure}

As described in the main text, by numerically solving Eqs.~(\ref{eq:fluidacceleration}), (\ref{eq:massconservation}) and (\ref{eq:entropyconservation}) on a grid of points, we established the dependence of the critical far density on the physical parameters of the problem: $L$, $\epsilon$, $M$, and $R$. Here, we give an illustration of this process. Fixing the starting radius to $R=20$ km and the average neutrino energy to $\epsilon=20$ MeV, we explore the relationship between luminosity and density at criticality. The results are shown in Fig.~\ref{fig:critical_numerical}, where the points show numerically determined critical parameters and the lines show the inferred power law fit. The quality of the power law fit given in Eq.~(\ref{eq:scaling_law_rho}) to the numerical data points is apparent. The two sets of points correspond to the values of $M_{PNS} = 1.4M_\odot$ and $M_{PNS} = 2.0M_\odot$.

For luminosity values below the critical points, the solution is subsonic, while above it is supersonic, with a termination shock. The power law can also be understood analytically, as  will be discussed in detail in \cite{paper2}.

\end{document}